\documentclass[aps,12pt,
		prd,
		reprint,
		onecolumn,
		superscriptaddress,
		shortbibliography,
		nofootinbib,
		floatfix,
		notitlepage
		]{revtex4-1}

\usepackage{amsmath,amssymb,amsfonts}

\usepackage{natbib}
\usepackage[T1]{fontenc}
\usepackage[latin1]{inputenc}

\usepackage[dvipsnames]{xcolor}

\usepackage{hyperref}
\hypersetup{colorlinks=true,citecolor=blue,urlcolor=Red}

\usepackage{mathrsfs}
\usepackage{bbm}
\usepackage{slashed}
\DeclareSymbolFont{rsfs}{U}{rsfs}{m}{n}
\DeclareSymbolFontAlphabet{\mathrsfs}{rsfs}
\usepackage[mathscr]{eucal}
\usepackage[normalem]{ulem}
\usepackage{verbatim}
\usepackage{bm}

\usepackage{graphicx}

\newcommand{\be}{\begin{equation}}
\newcommand{\ee}{\end{equation}}
\newcommand{\bi}{\begin{itemize}}
\newcommand{\ei}{\end{itemize}}
\newcommand{\bea}{\begin{eqnarray}}
\newcommand{\eea}{\end{eqnarray}}
\newcommand{\ud}{\mathrm{d}}		% roman d

\newcommand{\LCm}{{\scriptscriptstyle -}} %LC supersripts
\newcommand{\LCp}{{\scriptscriptstyle +}}
\newcommand{\LCpm}{{\scriptscriptstyle \pm}}
\newcommand{\LCmp}{{\scriptscriptstyle \mp}}

\newcommand{\LCperp}{{\scriptscriptstyle \perp}}

\newcommand{\ns}{\ensuremath{\slashed{n}}}
\newcommand{\as}{\ensuremath{\slashed{a}}}
\newcommand{\As}{\ensuremath{\slashed{A}}}

\newcommand{\ps}{\ensuremath{\slashed{p}}}

\DeclareMathOperator{\Ai}{Ai}
\DeclareMathOperator{\arcsinh}{arcsinh}

\usepackage{tikz}
\usetikzlibrary{decorations.markings}
\usetikzlibrary{positioning}
\usetikzlibrary{calc}
\usetikzlibrary{patterns}
\usetikzlibrary{decorations.pathmorphing, decorations.pathreplacing}

\tikzset{/pgf/decoration/.cd,
    number of sines/.initial=10,
    angle step/.initial=20,
}
\newdimen\tmpdimen
\pgfdeclaredecoration{complete sines}{initial}
{
    \state{initial}[
        width=+0pt,
        next state=move,
        persistent precomputation={
            \pgfmathparse{\pgfkeysvalueof{/pgf/decoration/angle step}}%
            \let\anglestep=\pgfmathresult%
            \let\currentangle=\pgfmathresult%
            \pgfmathsetlengthmacro{\pointsperanglestep}%
                {(\pgfdecoratedremainingdistance/\pgfkeysvalueof{/pgf/decoration/number of sines})/360*\anglestep}%
        }] {}
    \state{move}[width=+\pointsperanglestep, next state=draw]{
        \pgfpathmoveto{\pgfpointorigin}
    }
    \state{draw}[width=+\pointsperanglestep, switch if less than=1.25*\pointsperanglestep to final, % <- bit of a hack
        persistent postcomputation={
        \pgfmathparse{mod(\currentangle+\anglestep, 360)}%
        \let\currentangle=\pgfmathresult%
    }]{%
        \pgfmathsin{+\currentangle}%
        \tmpdimen=\pgfdecorationsegmentamplitude%
        \tmpdimen=\pgfmathresult\tmpdimen%
        \divide\tmpdimen by2\relax%
        \pgfpathlineto{\pgfqpoint{0pt}{\tmpdimen}}%
    }
    \state{final}{
        \ifdim\pgfdecoratedremainingdistance>0pt\relax
            \pgfpathlineto{\pgfpointdecoratedpathlast}
        \fi
   }
}
\tikzset{
    photon/.style={decorate, decoration={complete sines, number of sines = 8}, draw=black, thick},
    electron/.style={draw=black, thick},
    dressed/.style={draw=black, double, thick}
}
\usepackage{environ}
\NewEnviron{feynman}{\expandafter\begin{tikzpicture}[baseline={([yshift=-.5ex](0, 0))}]
    \BODY
    \end{tikzpicture}
}

\usepackage[caption=false,labelformat=simple,listofformat=subsimple]{subfig}

\newcommand{\subplot}[1]{\makeatletter%
    \if@twocolumn%
        \includegraphics[height=0.618\columnwidth]{#1}
    \else%
        \includegraphics[height=0.309\columnwidth]{#1}
    \fi
    \makeatother
}

\begin{document}
\title{High-intensity scaling in UV-modified QED}

\author{Robin Ekman}
\email{robin.ekman@plymouth.ac.uk}
\author{Tom Heinzl}
\email{theinzl@plymouth.ac.uk}
\author{Anton Ilderton}
\email{anton.ilderton@plymouth.ac.uk}

\affiliation{Centre for Mathematical Sciences, University of Plymouth, Plymouth, PL4 8AA, UK}

\begin{abstract}
    QED perturbation theory in a background field has been conjectured to break down for sufficiently high field intensity.
    The high-intensity behavior of a field theory is however intertwined with its high-energy (UV) behavior.
    Here we show that a UV modification of QED changes the high-intensity behaviour of observables.
    Specifically we study nonlinear Compton scattering in a constant crossed field in QED with an additional Pauli term.
    In the UV modified theory the cross section exhibits a faster power-law scaling with intensity than in ordinary QED.
\end{abstract}
\maketitle
\section{Introduction}

Strong electromagnetic fields can be realised using modern intense lasers~\cite{vulcan_site,ELI_SDRb}, made possible through chirped pulse amplification~\cite{Strickland:1985gxr}.  A crude model for strong laser fields is a background constant crossed field (CCF) with vanishing field invariants, $\mathcal{S} \equiv E^2 - B^2 =0$ and $\mathcal{P} \equiv \mathbf{E} \cdot \mathbf{B} = 0$, hence `null'. While this may seem a peculiar choice, it has been argued that any field configuration will appear as a CCF to a probe of sufficiently high energy~\cite{Nikishov:1964zza} (although there are caveats to this statement~\cite{Harvey:2014qla,DiPiazza:2017raw,Ilderton:2018nws}.) Given a probe, one can construct a nonvanishing invariant $\chi$, which is, roughly, the product of probe energy and CCF field strength; for an electron, $\chi = F_*/E_S$, that is the electric field $F_*$ seen by the electron in its rest frame, in units of the Sauter-Schwinger field strength, $E_S = m^2/e$ \cite{Sauter:1931zz,Schwinger:1951nm}.

It has been noted~\cite{Nikishov:1964zza} that the strong field asymptotics of loop diagrams in a background CCF scale with an effective coupling $g \equiv \alpha \chi^{2/3}$ --- see \cite{Fedotov:2016afw,Mironov:2020gbi} for a comprehensive discussion with an exhaustive list of references. This has been formalised into the \emph{Narozhny-Ritus conjecture}~\cite{Narozhnyi:1968,Ritus1,Narozhnyi:1979at,Narozhnyi:1980dc} that, in a CCF background, QED perturbation theory is an expansion in the effective coupling $g$. The conjecture thus implies that strong field perturbation theory in $g$ breaks down when $g \gtrsim 1$. See~\cite{Yakimenko:2018kih,Blackburn:2018tsn,DiPiazza:2019vwb} for experimental proposals on how to approach this regime.

The breakdown of perturbation theory at $g \gtrsim 1$ is  reminiscent of what happens in standard QED, where vacuum polarisation loops induce an effective expansion parameter at high energies, $g_0(q^2) \equiv (\alpha/3\pi) \ln q^2/m^2$, enhanced through large logarithms, with $q$ being the large momentum scale involved~\cite{Weinberg:1996kr,Schwartz:2013pla}. This implies the usual, resummed, running coupling, $\alpha(q^2) = \alpha \sum_n g_0^n = \alpha/(1 -g_0(q^2))$. Analogous logarithms appear at high field strengths where the effective expansion parameter becomes $g_0(\mathcal{S}) \equiv (\alpha/3\pi) \ln e(2|\mathcal{S}|)^{1/2}/m^2$ \cite{Migdal:1972sy,Ritus:1975pcc,Ritus:1977bis,Artimovich:1990qb,Kurilin:1999qc}.  As such, external field strength may be used to define the running of the coupling~\cite{Matinyan:1976mp,Dunne:2002ta}, providing the basis for the background field method of renormalisation~\cite{Abbott:1980hw,Abbott:1981ke}.  In the case of null fields, where $\mathcal{S} = 0$, the argument of the logarithm is instead essentially $\chi$~\cite{Artimovich:1990qb}. A dependence on $\ln \chi$, rather than $\alpha\chi^{2/3}$, has also been noted in the asymptotic limit of scattering amplitudes when high~$\chi$ is reached through high energy, in nonconstant fields~\cite{Podszus:2018hnz,Ilderton:2019vot}.

There is thus a close relationship between the strong-field behaviour of QED and its high-energy behaviour. Classically this is clear, as high intensities accelerate particles to high energies, but quantum mechanically the issue can be more subtle.
It has been argued~\cite{Migdal:1972sy} that in strong electric fields $E$, so $\mathcal{S} > 0$, the electron propagator becomes localised at an `electric length' scale of order $\ell_E \equiv 1/(eE)^{1/2}$, so that the argument of the strong-field logarithm is the ratio $\ell_E/\lambdabar_e$, $\lambdabar_e$ being the Compton length.
In a magnetic field $B$, so $\mathcal{S}<0$, the magnetic length $\ell_B \equiv 1/(eB)^{1/2}$ corresponds to the localisation of the lowest Landau level~\cite{Landau:1930,Heinzl:2007ca}. We thus see that strong external fields impact the ultraviolet (UV) behaviour of QED by providing a field dependent high-energy (small length) scale. 

Here we address the opposite question: how does a high-energy (UV) modification of QED affect its behaviour at high-intensities? One possible modification is the addition of a dimension-$5$ Pauli term $\sim \overline{\psi} \sigma F \psi$ to the action, coupling the fermion spin to the field.
Such a term is generated perturbatively at one loop in QED and defines the anomalous magnetic moment of the electron.
From an effective field theory point of view, a Pauli term corresponds to adding a lowest-order irrelevant interaction with the coupling suppressed by $1/\Lambda$, the scale where new physics should set in.
The addition of a Pauli term will therefore alter the UV behaviour of the theory, i.e., for momentum scales of order $\Lambda$. 
Indeed, it has been shown~\cite{Djukanovic:2017thn} that adding a Pauli term removes the QED Landau pole~\cite{Landau:1965nlt}. 
(See also \cite{Gockeler:1997dn} for lattice simulations and \cite{Gies:2004hy} for a renormalisation group analysis of this issue.)
For particular choices of the couplings $e$ and $\kappa$, the resulting theory may even become  asymptotically safe, hence UV complete~\cite{Gies:2020xuh}. 

We will show here that the presence of the Pauli term indeed changes the strong field behaviour of observables. We focus on a particular scattering process, namely nonlinear Compton scattering \cite{Nikishov:1964zza,Nikishov:1964zz,Brown:1964zzb,Goldman:1964,Seipt:2010ya} (reviewed in, e.g.~\cite{Seipt:2017ckc}), in a CCF background, as this is the case to which the NR conjecture most concretely applies.

The paper is organized as follows.
In Sect.~\ref{sec:pauli-volkov} we discuss the physics of the exact solutions to the Dirac-Pauli equation in a background plane wave.
In Sect.~\ref{sec:NLC} we use these solutions to calculate the probability of our chosen test process, nonlinear Compton scattering, specifying to the case of a CCF.
We derive the asymptotic scaling of the nonlinear Compton probability in Sect.~\ref{sec:scaling} and conclude in Sect.~\ref{sec:conclusions}.

\section{QED with a Pauli term}
\label{sec:pauli-volkov}

The Lagrangian of QED with an additional (dimension 5) Pauli term is
\begin{equation}
    \label{eq:action}
    \mathcal{L} = -\frac{1}{4} F^{\mu\nu}F_{\mu\nu} + \bar\Psi \left(i\slashed{\partial} -m - e \slashed{A} + \frac{\kappa}{\Lambda} \sigma^{\mu \nu} F_{\mu\nu} \right) \Psi \;,
\end{equation}
in which $\kappa$ is the \textit{dimensionless} Pauli coupling, $    \sigma^{\mu\nu} = \frac{i}{4} [ \gamma^\mu, \gamma^\nu]$, and $\Lambda$ is the cutoff scale as described in the introduction.
For convenience we redefine $\kappa \to \kappa \Lambda/m$.

We include also a background plane wave field $A_\text{ext}$, which shifts $A \to A + A_\text{ext} $ in the interaction terms of (\ref{eq:action})~\cite{Kibble:1965zza,Frantz,Gavrilov:1990qa,Ilderton:2017xbj}. (Later we will specialise to a CCF.) We are interested in the limit in which this field becomes strong, in a sense to be defined, but which amounts to the effective coupling between the background and matter eventually becoming (much) larger than unity. Hence we work in the Furry picture, in which all background-matter couplings are treated exactly, as part of the `free' theory, while interactions between the quantised fields are treated in perturbation theory as usual. As such, fermion propagators and external legs become `dressed' to all orders in the background field, in our case through both the QED vertex $A^\mu_\text{ext}\gamma_\mu$ and through the Pauli vertex $\sigma_{\mu \nu} F_\text{ext}^{\mu\nu}$.

\subsection{Intensity effects due to Pauli-term}

In the Furry picture, external fermion legs are given by incoming/outgoing solutions $\Psi$ to the Dirac equation in the chosen background, which now includes the Pauli term.
Writing $eA_\text{ext}^\mu = a^\mu$ the equation is 
\be\label{diracek}
    \left(i\slashed{\partial} -m - \slashed{a} + \frac{\kappa}{em} \sigma^{\mu \nu} f_{\mu\nu} \right) \Psi = 0 \;,
\ee
 with $f_{\mu\nu}$ the field strength of~$a_\mu$. The factor of~$1/e$ appears in the Pauli term just because we have absorbed~$e$ into the potential. Our chosen background is a (for the moment arbitrary) plane wave described by $a_\mu \equiv a_\mu(n\cdot x)$ in which $n^2=0$, $n\cdot a=0$, and $a$ has only spatial components. As is standard, we use lightfront coordinates (see~\cite{Brodsky:1997de,Heinzl:2000ht} for reviews) such that  $x \cdot n = x^\LCp =x^0 + x^3$, and the remaining spatial directions are $x^\LCm = x^0 - x^3$, $x^\LCperp = (x^1, x^2)$.
Momenta have components $p_\LCpm = (p_0 \pm p_3)/2$ and $p_\LCperp = (p_1, p_2)$. Note that $p^\LCpm = 2p_\LCmp$. The two nonzero components of $a_\mu$ are simply the integrals of the two electric field components of the wave, from $x^\LCp = - \infty$, see e.g.~\cite{Dinu:2012tj}.

 To understand the physics introduced by the Pauli term, it is useful to briefly review that of the Volkov wavefunctions, which are the solutions to (\ref{diracek}) with the Pauli term switched off.
(The high symmetry associated with plane waves guarantees the (super)integrability of both classical and quantum equations of motion~\cite{Heinzl:2017zsr,Heinzl:2017blq,Ansell:2018dro}.) For an incoming electron of initial momentum $p_\mu$ the appropriate Volkov solution is~\cite{volkov35}
 \begin{equation}
 \Psi_p(x)
    =
    \bigg(1 + \frac{\slashed{n}\slashed{a}(x^\LCp)}{2n\cdot p}\bigg) u_p\,
    \exp\bigg[
        -ip\cdot x - i\int\limits^{x^\LCp} \frac{2p\cdot a - a\cdot a}{2n\cdot p}
    \bigg] \;.
    \label{eq:volkov}
\end{equation}
The current of these wavefunctions recovers the classical, on-shell, kinematic momentum $\pi_\mu$ of an electron in a plane wave background, i.e.~$\bar\Psi_p(x) \gamma_\mu \Psi_p(x)/2 = \pi_\mu(x)$ with
\begin{equation}
\label{PiDef}
    \pi_\mu(x^\LCp) = p_\mu - a_\mu(x^\LCp) + n_\mu \frac{2 a(x^\LCp)\cdot p - a(x^\LCp)^2}{2 n \cdot p} \;.
\end{equation}
However, while interaction with a plane wave can change electron momentum, it cannot change the quantum spin state of the electron~\cite{Ilderton:2020gno}. To see this, first write the spin structure in (\ref{eq:volkov}) as
\begin{equation}
    u_\pi(x^\LCp) = \bigg(1 + \frac{\ns \as(x^\LCp)}{2 n \cdot p} \bigg) u_p \;.
    \label{eq:u-pi}
\end{equation}
It is easily checked that $\slashed{\pi} u_\pi =  mu_\pi$, hence $u_\pi$ is a `free' spinor for the (time-dependent) on-shell momentum $\pi_\mu$ and (\ref{eq:u-pi}) is not a trick of notation.
It is convenient to use a lightfront helicity basis~\cite{Chiu:2017ycx}, in which the spinors are eigenstates of the lightfront helicity operator  $L_p = 2h_p \cdot W/m$, that is the Pauli-Lubanski vector $W_\mu$ contracted with $h_p^\mu$ where~\cite{Itzykson:1980rh,Chiu:2017ycx},
\begin{equation}
    h^\mu_p = \frac{p^\mu}{m} - \frac{m}{n\cdot p} n^\mu \; , 
    \qquad
    L_p = -\frac{1}{m} \epsilon_{\mu\nu\alpha\beta} h_p^\mu p^\nu \sigma^{\alpha\beta}
    =
    \frac{1}{m} \gamma^5 \slashed{h}_p \ps \;,
\end{equation}
in which $p$ is understood as the momentum of the state acted on.
The eigenvalues of $L_p$ are $\pm 1$, and explicit forms of the corresponding eigenspinors $u_{p\LCpm}$ are given in Appendix~\ref{app:basis}.
(For discussions of other relativistic spin operators see~\cite{Chiu:2017ycx,Aleksandrov:2020xop}.) It is then easy to verify that
\begin{equation}
    L_\pi u_{\pi\LCpm}
    \equiv L_\pi \bigg(1 + \frac{\ns \as(x^\LCp)}{2 n \cdot p} \bigg) u_{p\LCpm}
    = \left(1 + \frac{\ns \as(x^\LCp)}{2 n \cdot p} \right) L_p u_{p \LCpm}
    = \pm u_{\pi\LCpm}
\;,
\end{equation}
so that helicity is preserved in the plane wave background. As such the Volkov solutions do not describe helicity-flip transitions, for which either loop corrections or photon emission is needed~\cite{Ilderton:2020gno}.

We now return to the Dirac equation with Pauli term~\eqref{diracek}.
For a linearly polarised field (as we adopt below) $A_\text{ext}^\mu = A(x^\LCp) \epsilon^\mu$ with $\epsilon^2=-1$, $n\cdot \epsilon=0$, the solution $\psi_p(x)$ to \eqref{diracek} is given by replacing, in~\eqref{eq:volkov},  $u_p \mapsto U_p$ where~\cite{Chakrabarti:1968zz}
\begin{align}
    U_p
    =
    \frac{\ps + m}{2n \cdot p} \ns \exp(-\kappa \As / m)
    u_p
    =
    \frac{\ps + m}{2n \cdot p} \ns\big(
        \cos (\kappa A / m) - \sin (\kappa A / m) \slashed{\epsilon}
    \big) u_p
    \;.
    \label{eq:chakrabarti}
\end{align}
It can be checked that the current of these solutions is the same as that of the Volkov solutions. The physical content of the literature result (\ref{eq:chakrabarti}) is made transparent using our helicity basis. Using either the explicit representation given in Appendix~\ref{app:basis}, or the defining properties of the states, one can show that 
\begin{equation}
    \label{useful}
    \frac{\ps + m}{2n \cdot p} \ns \,  u_{p\LCpm} = u_{p\LCpm} \;,
    \qquad 
    \frac{\ps + m}{2n \cdot p} \ns \slashed{\epsilon} u_{p\LCpm} = \pm u_{p\LCmp} \;.
\end{equation}
Hence the `cosine' term in (\ref{eq:chakrabarti}) preserves the helicity of the state, while the `sine' term flips it. The extra spin structure due to the Pauli term therefore amounts to a rotation matrix in helicity space:
\begin{equation}
    \begin{pmatrix}
        U_{p\LCp} \\
        U_{p\LCm}
    \end{pmatrix}
    =
    \begin{pmatrix}
        \cos \kappa A/m & -\sin \kappa A/m \\
        \sin \kappa A/m & \cos \kappa A/m
    \end{pmatrix}
    \begin{pmatrix}
        u_{p\LCp} \\
        u_{p\LCm}
    \end{pmatrix}
    .
    \label{eq:rot-mat}
\end{equation}
From \eqref{eq:u-pi}, this result extends immediately to the spinors $u_{\pi\LCpm}$.
Physically, the Pauli term therefore causes an initial helicity eigenstate to evolve into a superposition of helicity states as the particle propagates through the plane wave.
In standard QED, such effects arise, implicitly and usually considered only perturbatively, through one-loop corrections, which generate a Pauli term and thus the anomalous magnetic moment of the electron.

\section{Nonlinear Compton scattering}
\label{sec:NLC}

In this section we calculate the probability of nonlinear Compton scattering~\cite{Nikishov:1964zza,Nikishov:1964zz,Brown:1964zzb,Goldman:1964}.
In terms of the Volkov wavefunctions above, this is a $1\to 2$ process in which a dressed electron emits a photon. Physically, the electron is incident on the plane wave background, is accelerated, and emits radiation. If the process is calculated perturbatively in the background, one sees that the leading order contribution is ordinary $2\to 2$ Compton scattering, with the incoming photon sourced from the background. Treating the background exactly, through the Volkov wavefunctions, means allowing all numbers of background photons to participate in the process, hence the name nonlinear Compton scattering; see~\cite{Seipt:2017ckc} for a recent review. We focus here on the long wavelength limit, i.e.\ we treat the background as a CCF, as this is the case to which the NR conjecture most concretely applies.

The field is described by the potential  $a_\mu = eE\epsilon_\mu x^\LCp$, where, without loss of generality, $\epsilon\cdot x = x^1$. For an electron of momentum $p_\mu$, observables in a CCF depend on the `quantum nonlinearity parameter' $\chi$ defined by
\be
    \chi = \sqrt{\frac{-(eF\cdot p)^2}{m^6}} = \frac{n\cdot p}{m} \frac{|E|}{E_S} \;.
\ee
We focus on the CCF case as there is mounting evidence that universal power-law scaling at large $\chi$ is absent for non-constant fields~\cite{Podszus:2018hnz,Ilderton:2019vot,Ilderton:2019kqp,Torgrimsson:2020wlz,Edwards:2020npu}. In a CCF, the electron self-energy loop shows the typical behaviour stated by the Narozhny-Ritus conjecture, which is a large-$\chi$ scaling with $\alpha \chi^{2/3}$. The same scaling is inherited, via the optical theorem, by nonlinear Compton scattering at tree level. Another reason to focus on CCFs is their common use in laser-plasma simulation codes~\cite{Elkina:2010up,Gonoskov:2014mda} in the form of the locally constant field approximation (LCFA)~\cite{Ritus1985a}.

Let $\psi_{p\LCpm}(x)$ be the helicity eigenstates given by (\ref{eq:chakrabarti}) and (\ref{eq:volkov}). The nonlinear Compton scattering amplitude is
\begin{align}
    \mathcal{M}
    & =
    \int\!\ud^4 x\,
    \overline{\psi}_{p' s'}(x) e^{i\ell\cdot x}  { \varepsilon}_\mu \bigg(
        -ie\gamma^\mu + i\frac{\kappa}{m} \gamma^\mu\slashed{\ell}
    \bigg) 
    \psi_{p s}(x)
    \\
    & =
    \def\r{1}
    \def\q{40}
    \begin{feynman}
        \draw [dressed] (-\r,0) --
            node[at start, left] {$p$}
            node[at end, right] {$p'$}
        (\r, 0);
        \draw [photon, decoration={number of sines=5}] (0, 0.01) --
            node[at end, right] {$\ell$}
        ++ (\q:\r);
    \end{feynman}
    +
    \begin{feynman}
        \draw [dressed] (-\r,0) --
        node[at start, left] {$p$}
        node[at end, right] {$p'$}
        (\r, 0);
        \draw [photon, decoration={number of sines=5}] (0, 0.01) --
        node[at end, right] {$\ell$}
        ++ (\q:\r);
        \filldraw [fill=white] (0,0) circle (.1);
    \end{feynman}
    \label{eq:amp-diagram}
    \;,
\end{align}
in which $\ell_\mu$ and $\varepsilon_\mu$ are the momentum and polarisation vector of the outgoing photon, respectively.  On the lower line the left hand diagram represents the standard QED vertex $-ie\gamma^\mu$ and the right hand diagram the Pauli vertex $i \kappa \gamma^\mu \slashed{\ell}/m$. We consider the nonlinear Compton probability summed and averaged over spins and polarisations. While the Pauli term simply induces a rotation in helicity space, the spin-summed probability can still exhibit nontrivial effects because the rotation is (lightfront) \emph{time dependent}, and therefore cannot simply be factored out of the amplitude. By standard arguments, see e.g.~\cite{Seipt:2017ckc}, the probability can be brought to the form
\begin{equation}
    \label{Prob1}
    \mathbb{P}
    =
    \frac{m^2}{4 n\cdot p }
    \int \frac{\ud^2 \ell_\perp}{(2 \pi)^3}
    \int_0^\infty \frac{\ud s}{s}
    \int
    \frac{\ud x^\LCp \, \ud y^\LCp}{n \cdot (p - \ell)}
    \,
    \Upsilon
    \exp\left\{\frac{i}{n \cdot (p - \ell)} \int^{x^+}_{y^+} \ell \cdot \pi_p(z) \, \ud z \right\}
\end{equation}
in which the momentum $p'_\mu$ of the outgoing electron is eliminated using conservation of momentum in the field, and $s := n\cdot \ell/n\cdot p = \ell_\LCm / p_\LCm$ is the lightfront momentum fraction of the emitted photon. The modifications with respect to standard QED are encoded in the quantity $\Upsilon$, which contains the (time-dependent) Dirac structure coming from the vertices and the Volkov-Pauli solution \eqref{eq:volkov} and \eqref{eq:chakrabarti}: writing $\pi'$ for the classical momentum related to $p'$ as $\pi$ is to $p$ in (\ref{PiDef}),
\begin{equation}
    \Upsilon=
    \frac{1}{4m^2} \sum\limits_{s,s',\varepsilon} \bigg[
        {\bar U}_{\pi's'}(x^\LCp) \bigg(
            e \slashed{\varepsilon}
            - \frac{\kappa}{m} \slashed{\varepsilon}\slashed{\ell}
        \bigg) U_{\pi s}(x^\LCp)
    \bigg]
    \bigg[{\bar U}_{\pi s}(y^\LCp)\bigg(
            e\slashed{\varepsilon}^\dagger
            - \frac{\kappa}{m} \slashed{\ell}\slashed{\varepsilon}^\dagger
    \bigg) {U}_{\pi' s'}(y^\LCp) \bigg]
    \;.
\end{equation}
Defining $ \Delta := \frac{\kappa}{m} E\big( x^\LCp -y^\LCp\big)$, a rescaled time difference, the fermion spin sums are evaluated using~\eqref{eq:chakrabarti}, which yields
\begin{equation}
    \sum_s U_{p s}(x^\LCp) \overline{U}_{p s}(y^\LCp)
    =
    (\ps + m) \cos \Delta
    +
    \frac{ \ps + m }{2 n \cdot p} \slashed{\epsilon} \ns (\ps + m)
    \sin \Delta
    \;.
    \label{eq:spin-sum}
\end{equation}
In contrast to standard QED ($\kappa=0$), the new Dirac structure introduces trigonometric functions of lightfront time, seen in (\ref{eq:spin-sum}), into $\Upsilon$, under the integrals of (\ref{Prob1}). This will have considerable impact on the evaluation of the emitted photon momentum integrals.

Performing the Gaussian integral over $\ell_\perp$ in \eqref{Prob1} eliminates any term linear in $\ell_\perp$ from $\Upsilon$ and yields a multiplicative factor for any term quadratic in $\ell_\perp$. Now, some terms in $\Upsilon$ are proportional to $\ell \cdot \pi_p(x^\LCp)$, $\ell \cdot \pi_p(y^\LCp)$ or products thereof, like the exponent of (\ref{Prob1}). Normally, for $\kappa = 0$, these terms are total derivatives when integrating over $x^\LCp$ or $y^\LCp$ and can be discarded~\cite{Boca:2009zz,Dinu:2012tj,Ilderton:2020rgk}. However, due to the presence of the trigonometric functions in \eqref{eq:spin-sum}, stemming from the Pauli term, some contributions remain after integration by parts. To illustrate, one such term is proportional to (the unwritten exponent is exactly as in (\ref{Prob1})),
\begin{equation}
    \label{eq:ibp-example}
    \begin{split}
        \ell \cdot ( \pi_p(x^\LCp) + \pi_p(y^\LCp)) \cos^2 (\Delta)
        \exp[\cdots]
    & =
    -i n\cdot(p-\ell) \cos^2 (\Delta) (\partial_{x^+} - \partial_{y^+})
        \exp[\cdots]
        \\
    & \mapsto -i \frac{2\kappa E}{m} n\cdot(p-\ell) \sin(2\Delta)
        \exp[\cdots] \;.
    \end{split}
\end{equation}
The following standard change to average and relative phase variables, 
\begin{equation}
    \varphi = \frac{1}{2}(x^\LCp + y^\LCp)
    \,,
    \qquad
    \theta = \frac{m^2 \chi \sqrt{z} }{2(n \cdot p) } (x^\LCp - y^\LCp)
    \,,
    \qquad
    z
    =
    \left( \frac{s}{\chi(1-s)} \right)^{2/3}
    \,,
    \label{eq:z-def}
\end{equation}
brings the probability into `Airy form',
\begin{equation}
    \mathbb{P}
    =
    \frac{m^2}{4 i p_-}
    \int_0^1 \!\ud s \,
    \int\!
    \frac{\ud\varphi}{2\pi}
    \frac{\ud \theta}{\theta}
    \,
    \tilde{\Upsilon}
    \,
    \exp i \left(z \theta + \frac{\theta^3}{3}\right)
    \;,
    \label{eq:almost-final}
\end{equation}
in which $\tilde{\Upsilon}$ is a sum of powers of $\theta$, each carrying a trigonometric factor; for example, the term in (\ref{eq:ibp-example}) contains the factor $\sin(k \theta/\sqrt{z})$ in which we define
\begin{equation}
    k = \frac{4\kappa}{e} \;.
\end{equation}
Note that the argument of such trigonometric functions depends on both the Pauli coupling and (through $z$) on $\chi$.
If not for the trigonometric factors in $\tilde{\Upsilon}$, the $\theta$-integrals in the probability could be expressed in terms of Airy functions using the integral representations (with an $i\varepsilon$ prescription understood)
\begin{align}
    \frac{i^n}{2 \pi} \int \!\ud \theta \, \theta^n e^{i \theta z + i \theta^3/3}
    & =
    \frac{\ud^n \Ai}{\ud z^n}(z)
    \\
    -\frac{i^n}{2 \pi} \int\!\ud \theta \, \theta^{-n} e^{i \theta z + i \theta^3/3}
    & =
    \int_z^\infty\!\ud z_1 \cdots \int_{z_{n-1}}^\infty \ud z_n \Ai(z_n) 
    =: \Ai_n(z)
    .
\end{align}
However, by writing the trigonometric functions in exponential form, we see that their effect is just that of a finite difference operator acting on the Airy functions, e.g.,
\begin{equation}
    -\frac{1}{2 \pi i}  \int \frac{d\theta}{\theta}
    \left(
        e^{i  (z + k/\sqrt{z})\theta + i\theta^3/3   }
        - e^{i (z - k/\sqrt{z})\theta + i\theta^3/3   }
    \right)
    =
    \Ai_1(z + k/\sqrt{z}) - \Ai_1(z - k/\sqrt{z})
\end{equation}
To make our expressions more manageable, we introduce the following shorthand notation for this operator, and others that appear,
\begin{align}
    \mathcal{D}_\pm f(z)
    & :=
    \frac{1}{2} \left( f(z + k/\sqrt{z}) \pm f(z - k/\sqrt{z}) \right)
    \\
    \mathcal{C} f(z)
    & :=
    \frac{1}{2} f(z) + \frac{1}{4} \left(f(z+ k/\sqrt{z}) + f(z - k/\sqrt{z}) \right)
    \\
    \mathcal{S} f(z)
    & :=
    \frac{1}{2} f(z) - \frac{1}{4} \left(f(z +k/\sqrt{z}) + f(z - k/\sqrt{z}) \right)
\end{align}
which ultimately arise from terms containing $\cos 2 \Delta, \sin 2 \Delta, \cos^2 \Delta, \sin^2 \Delta$, respectively -- as such, the operators obey the same relational identities as the trigonometric functions themselves.

The final result for the probability has three parts, proportional to $e^2, e \kappa$  and $\kappa^2$, respectively,
\begin{equation}
\label{tre-termer}
    \mathbb{P} =
    \frac{m^2}{4 \pi n\cdot p} \int \! \ud\varphi \int_0^1\!\ud s\,
    \left( e^2 \frac{\ud\mathbb{P}^{ee}}{\ud s}
    + e \kappa \frac{\ud\mathbb{P}^{e\kappa}}{\ud s}
    + \kappa^2 \frac{\ud\mathbb{P}^{\kappa\kappa}}{\ud s} \right) \;,
\end{equation}
corresponding to the three terms in $|\mathcal{M}|^2$,
\begin{equation}
    \label{eq:m-sq-expanded}
    \def\r{1}
    \def\q{40}
    |\mathcal{M}|^2
    =
    \Big|\begin{feynman}
        \draw [dressed] (-\r,0) --
        (\r, 0);
        \draw [photon, decoration={number of sines=5}] (0, 0.01) --
        ++ (\q:\r);
    \end{feynman}
    \Big|^2
    +2 \operatorname{Re} \Big(
    \begin{feynman}
        \draw [dressed] (-\r,0) --
        (\r, 0);
        \draw [photon, decoration={number of sines=5}] (0, 0.01) --
        ++ (\q:\r);
    \end{feynman} \times \begin{feynman}
        \draw [dressed] (-\r,0) --
        (\r, 0);
        \draw [photon, decoration={number of sines=5}] (0, 0.01) --
        ++ (\q:\r);
        \filldraw [fill=white] (0,0) circle (.1);
    \end{feynman}^\dagger
    \Big)
    +
    \Big|\begin{feynman}
        \draw [dressed] (-\r,0) --
        (\r, 0);
        \draw [photon, decoration={number of sines=5}] (0, 0.01) --
        ++ (\q:\r);
        \filldraw [fill=white] (0,0) circle (.1);
    \end{feynman}
    \Big|^2
    \; .
\end{equation}
Explicitly, we find
\begin{eqnarray}
    \frac{\ud\mathbb{P}^{ee}}{\ud s}
    &=&
    -\Ai_1(z)
    - \frac{2}{z} \Ai'(z)
    - s \chi \sqrt{z} \, \mathcal{C} \Ai'(z)
    - s z \, \mathcal{D}_- \Ai(z) \; , 
    \label{eq:P-ee} \\[10pt]
    \frac{d\mathbb{P}^{e\kappa}}{ds}
    &=&
    4s \chi \mathcal{D}_- \Ai''(z)
    + 4s \chi z^{1/2}
         \left(1 + \mathcal{S}\right) \Ai'(z)
    - 6\frac{\kappa  \chi}{e} s z^{-1/2} \mathcal{D}_+ \Ai(z)
    + 4 s \chi z^{3/2}
     \mathcal{S} \Ai_1(z)
    - 2 s \chi \sqrt{z} \mathcal{S} \Ai_2(z) \; , \label{eq:P-ek}\\[10pt]
    \frac{\ud \mathbb{P}^{\kappa\kappa}}{\ud s}
    &=&
    - \frac{8s \chi}{z^{1/2}} \mathcal{C}\Ai'''(z)
    - \frac{8s \chi}{z^{1/2}} \mathcal{D}_- \Ai''(z)
    +   \left( \frac{16\kappa  \chi}{e z} \mathcal{D}_-
            +
        s \chi z^{1/2} (-2 + 3 \mathcal{D}_+) \right)\Ai'(z)
    \nonumber \\
    &+&  \left(
            \frac{24 \kappa  \chi}{e z^{1/2}} \mathcal{D}_+
            +
            \frac{24 \chi s}{z^{1/2}} \mathcal{C} + 4s \chi z \mathcal{D}_-
    \right) \Ai(z)
    - 4s \chi
        \left(z^{3/2} \mathcal{S} + \Big(2 + \frac{2\kappa }{e} \Big) \mathcal{D}_- \right)
    \Ai_1(z)
    + 2s\chi z^{1/2} \mathcal{S} \Ai_2(z)
    .
    \label{eq:P-kk}
\end{eqnarray}
The usual CCF result is recovered from (\ref{eq:P-ee}) by setting $\kappa = 0$, upon which $\mathcal{C} \mapsto 1$ and $\mathcal{D}_- \mapsto 0$. The shifts of $k/\sqrt{z}$ in the arguments of the Airy functions encode a change in the photon spectrum, relative to ordinary QED, through the $s$-dependence in $z$. Further, one sees in the same combination $k/\sqrt{z} \sim \kappa \chi^{1/3}/e$ an explicit interplay between the high-energy modification of the theory and the intensity dependence. The high precision to which the anomalous magnetic moment of the electron is known would, phenomenologically, limit $\kappa$ (equivalently $k$) to small values. This means that there will be, typically, only minor phenomenological effects from the additional terms in (\ref{eq:P-ee})--(\ref{eq:P-kk}). However, our interest is in the regime of extreme field strengths, and we will now show that the high-intensity behaviour of the probability above is dominated by terms coming from the Pauli interaction.
Viewed another way, we will show that strong fields essentially enhance $\kappa$.

\section{High-intensity behaviour}
\label{sec:scaling}
In this section we establish the asymptotic strong field behaviour of the emission probability (\ref{tre-termer}).
To extract this from (\ref{eq:P-ee})--(\ref{eq:P-kk}) we must first understand integrals of the type
\begin{equation}
    \int_0^1 \ud s \, s^a z^p \mathcal{A}(z + \lambda/\sqrt{z})
    \qquad
    a = 0, 1
    \;,
    \label{eq:integral-form}
\end{equation}
where $\mathcal{A}$ is $\Ai$ or one of its derivatives or integrals, and $\lambda$ is either $\pm k$ or $0$; the latter is needed because even some of the $\kappa$-dependent terms of (\ref{eq:P-ee})--(\ref{eq:P-kk}) have zero shift of the Airy function.
After a change of variables $s\to t$, where
\begin{equation}
 s = \frac{\chi t}{1 + \chi t} \;,
\end{equation}
the integrals (\ref{eq:integral-form}) take the form,
\begin{equation}
    J = \int_0^\infty \ud t \, \frac{\chi^{a+1} t^b}{(1 + \chi t)^{a +2 }} \mathcal{A}(z + \lambda/\sqrt{z} )
    \qquad
    b = a + 2p/3
    ,
    \label{eq:integral-form-t}
\end{equation}
where now $z = t^{2/3}$ and all $\chi$-dependence has been moved outside $\mathcal{A}$. We first consider $\lambda = 0$.
In order to obtain the asymptotic behaviour of our integrals we will need to distinguish between `small' and `large' $t$; thus we introduce a cutoff $L$, below which
$\mathcal{A}(z)$ can be approximated by a Taylor-MacLaurin series, its order $N$ determined by the desired accuracy of approximation.
This divides the integral as
\begin{equation}
    J \simeq J_0 + J_1
    :=
    \int_0^L \ud t \, \frac{\chi^{a+1} t^b}{(1 + \chi t)^{a+2}} \sum_{j = 0}^N \frac{a_j t^{2j/3}}{j!}
    +
    \int_L^\infty \ud t \, \frac{\chi^{a+1} t^b}{(1 + \chi t)^{a+2}} \mathcal{A}(z) \;
    ,
    \label{eq:maclaurin}
\end{equation}
For $\chi$ large enough, the $1$ in the denominator is negligible in the second term ($t > L$), so that $J_1$ is $O(\chi^{-1})$.
By simply performing the integral (change variables to $t' = 1 + \chi t$ and use the binomial series), one finds
\begin{equation}
    \int_0^L \ud t \, \frac{t^\beta}{(1 + \chi t)^{a+2}}
    =
    \begin{cases}{}
        O(\chi^{-1-\beta}) \; , & \beta < a + 1 \; , \\
        O( \chi^{-a-2} \ln \chi ) \; , & \beta = a + 1 \; , \\
        O( \chi^{-a-2} ) \; , & \beta > a + 1 \; .
    \end{cases}
    \label{eq:lower-integral-scaling}
\end{equation}
Since the scaling in $\chi$ decreases with $\beta$ the dominant contribution to $J_0$ comes from the leading $j = 0$ term of the Taylor-MacLaurin series, and the scaling of $J$ can now be read off from \eqref{eq:lower-integral-scaling} with $\beta = b = a + 2p/3$.

\begin{figure}[t!]
    \makeatletter%
    \if@twocolumn%
        \def\figwidth{0.618}
    \else%
        \def\figwidth{0.307}
    \fi
    \makeatother
    \begin{tikzpicture}[x = \figwidth*\columnwidth, y = \figwidth*\columnwidth]
        \def\L{1}
        \def\r{0.5}
        \def\q{0.2}
        \draw [<->, thick] (0, .6\L) |-
            node[at start, right] {$\operatorname{Im} u$}
            node[at end, below] {$\operatorname{Re} u$}
            (\L, 0);
        \node [
                inner sep=0pt, minimum size=2mm, circle, fill=black,
                label={left:$i\sqrt{\chi}$}
            ] (pole) at (0, .5) {};
        \node [
                inner sep=0pt, minimum size=2mm,
                label={below:$\varepsilon^{-1/2}$}
            ] (corner) at (\q, 0) {};
        \draw [dashed, postaction=decorate,
            decoration={markings,
                mark=between positions 0.2 and 0.8 step 0.3 with {{\arrow[black,line width=0.2mm]{>}};}
            } ] ($(corner) + (0, 0.5mm)$) -- ++ (\r, 0) arc (0:90:\r) -- cycle;
    \end{tikzpicture}
    \caption{\label{fig:contour}
        The upper contour choice for the integration in (\ref{eq:contour-integral}), to be complemented by its mirror image below the real axis.
        The mark on the imaginary axis indicates a pole of the integrand.
    }
\end{figure}
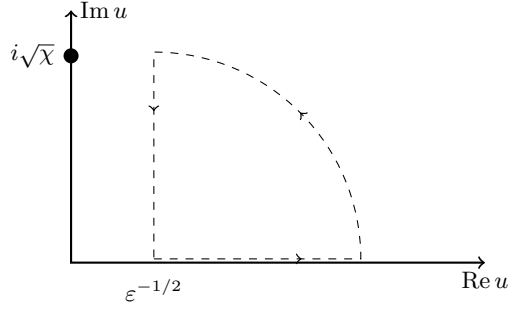

We now turn to the case $\lambda \neq 0$. Here we can divide the integration as
\begin{equation}
    J \simeq J_0 + J_1
    :=
    \int_0^\varepsilon \ud t \, \frac{\chi^{a+1} t^b}{(1 + \chi t)^{a+2}} \mathcal{A}(\lambda/\sqrt{z} ) +
    \int_\varepsilon^\infty \ud t \, \frac{\chi^{a+1} t^b}{(1 + \chi t)^{a+2}} \mathcal{A}(z + \lambda/\sqrt{z} ) \; ,
\end{equation}
the error in the argument of $\mathcal{A}$ being less than $\varepsilon$ in the first term ($t < \varepsilon$). By the same reasoning as before, $J_1 = O(\chi^{-1})$. If $\lambda > 0$, an asymptotic expansion for $\mathcal{A}$ around $+\infty$ can be used, viz.,
\begin{equation}
    J_0 \sim \int_0^\varepsilon \ud t \, \frac{\chi^{a+1} t^{b'}}{(1 + \chi t)^{a+2}} e^{ - \frac{2}{3}\lambda^{3/2} t^{-1/2}}
    \label{eq:airy-exp}
\end{equation}
where $b' = b + \{-\frac{1}{6}, \frac{1}{6}, -\frac{1}{2} \}$ when $\mathcal{A}=\{\Ai, \Ai', \Ai_1\}$.
While the integration extends down to $0$, if $\chi t \gg 1$ does not hold, then the log of the integrand is $O(-\sqrt{\chi})$, so non-negligible contributions come only from $t \gg 1/\chi$, hence $J_0$ is again $O(\chi^{-1})$. If $\lambda < 0$ and $\mathcal{A}$ is $\Ai$ or one of its derivatives, the asymptotic expansion around $-\infty$ is
\begin{equation}
    J_0 \sim \int_0^\varepsilon \ud t \, \frac{\chi^{a+1} t^{b'}}{(1 + \chi t)^{a+2}}
    \left\{ \begin{matrix} \cos \\ \sin \end{matrix}\right\} \left(
        \frac{2}{3} |\lambda|^{3/2} t^{-1/2}
    \right)
    =
    \int\limits_{\varepsilon^{-1/2}}^\infty \ud u \, \frac{\chi^{a+1} u^{2a - 2b' + 1}}{(u^2 + \chi)^{a+2}}
    \left\{ \begin{matrix} \cos \\ \sin \end{matrix}\right\} \left(
        \frac{2}{3} |\lambda|^{3/2} u
    \right)
    \label{eq:contour-integral}
    \; .
\end{equation}
The integral over $u := t^{-1/2}$ can be estimated by writing the integrand in exponential form and closing the contour through a quarter circle in the first or fourth quadrant to guarantee convergence, see Fig.~\ref{fig:contour}.
As $b' > -1$, the integral along the arc does not contribute in the asymptotic limit. (The integrand has poles at $\pm i \sqrt{\chi}$, but these are outside the contour.
In any case, the logs of the residues go like $-\sqrt{\chi}$, and so would yield exponentially suppressed contributions at large $\chi$.).
Now, on the vertical segments of the contour, the integrand goes like $e^{-|u|}$, so contributions outside $u \ll \chi$ are exponentially suppressed.
Thus, we again find $J_0 = O(\chi^{-1})$.

On the other hand, if $\mathcal{A}$ is $\Ai_1$, the asymptotic expansion is, for $C$ a constant,
\begin{equation}
    J_0 \sim \chi^{a+1} \int_0^\varepsilon \ud t \, \frac{t^b - C t^{b'} \cos( \cdots )}{(1 + \chi t)^{a+2}}  \,
\end{equation}
and the scaling of the first term can be read off from \eqref{eq:lower-integral-scaling}, while the second is $O(\chi^{-1})$ by the previous paragraph.
Finally, the Airy differential equation implies that $\Ai_2(x) = -x\Ai_1(x) - \Ai'(x)$, reducing terms with $\Ai_2$ to previously treated cases.

\begin{figure}[tp]
    \centering
    \subfloat[$f = \chi s \sqrt{z} \Ai'(z + \lambda/\sqrt{z} )$]{%
        \subplot{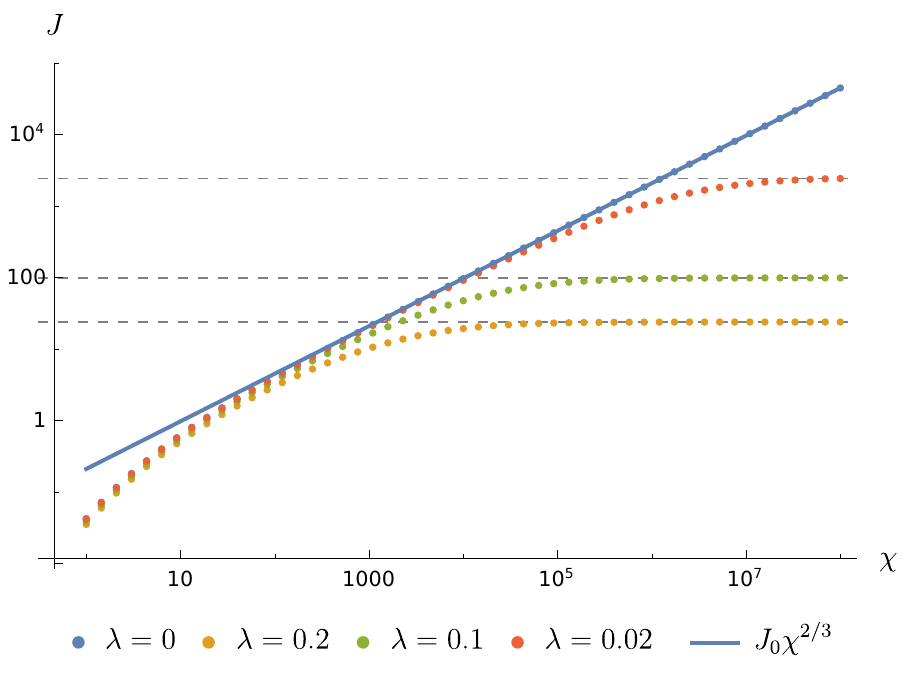}
    }
    \subfloat[$f = \chi s \sqrt{z} \Ai'(z + \lambda/\sqrt{z} )$, $\arcsinh$ scale on the vertical axis
    \label{subfig:ee-neg}]{%
        \subplot{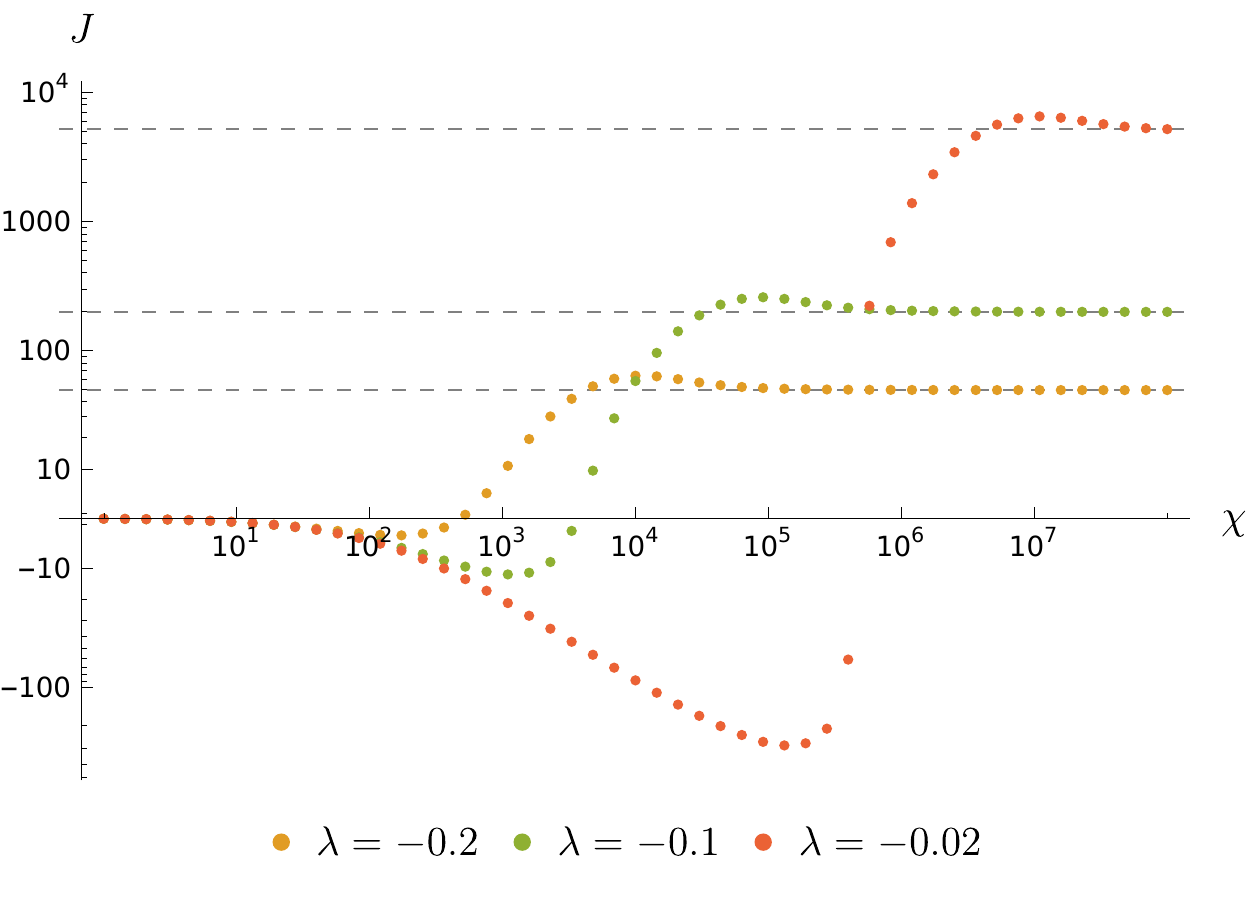}
    }\\
    \subfloat[$f = s \chi z^{1/2} \Ai_2(z + \lambda/\sqrt{z}) $]{%
        \subplot{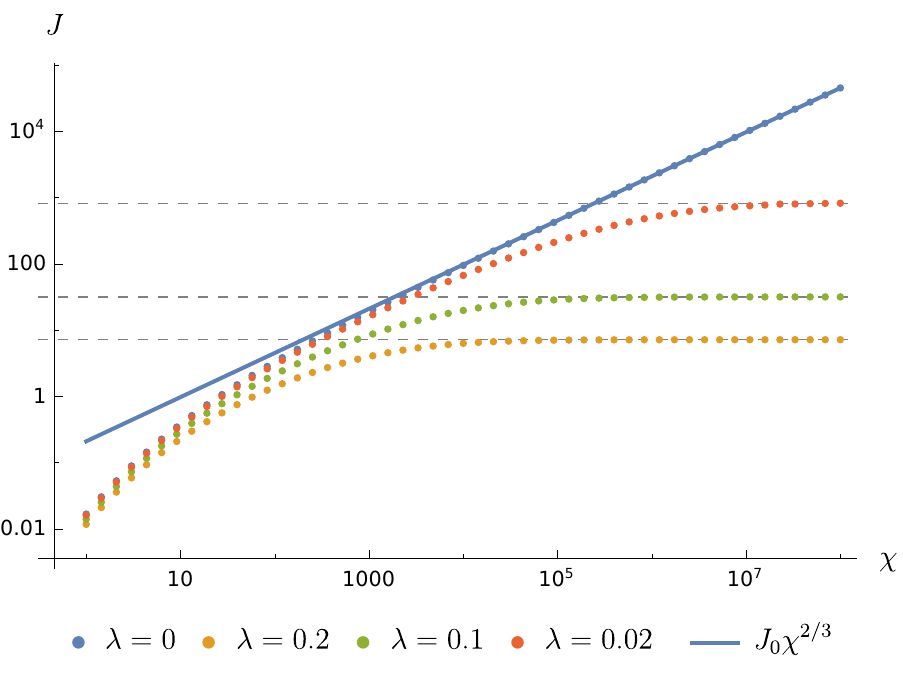}
    }
    \subfloat[$f = s \chi z^{1/2} \Ai_2(z + \lambda/\sqrt{z}) $]{%
        \subplot{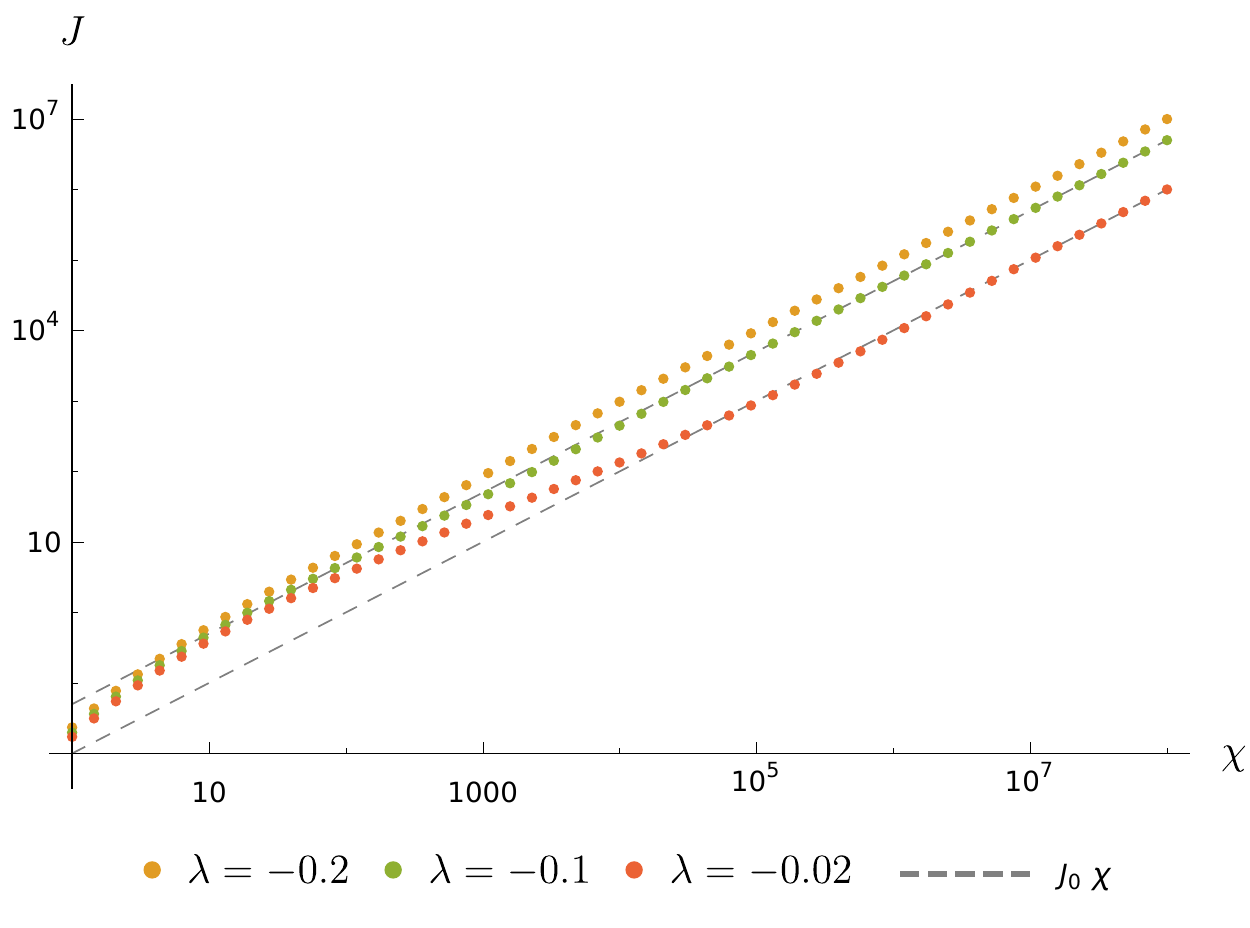}
    }\\
    \subfloat[$ f=\chi s z^{-1/2} \Ai(z + \lambda/\sqrt{z}) $\label{fig:incorrect-scaling}]{%
        \subplot{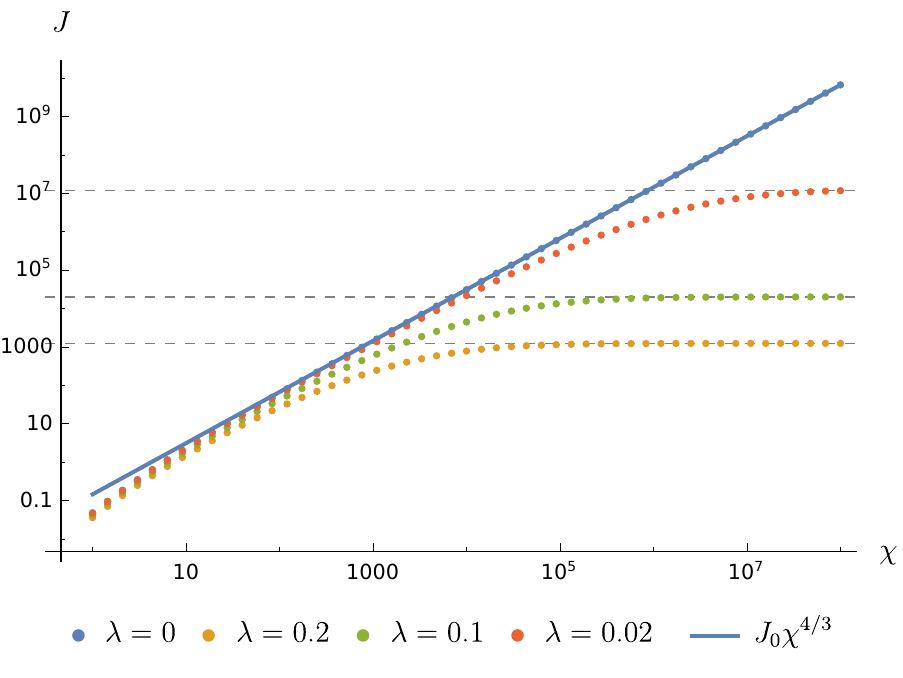}
    }
    \subfloat[$ f=\chi s z^{-1/2} \Ai(z + \lambda/\sqrt{z}) $, $\arcsinh$ scale on the vertical axis\label{subfig:kk-neg}]{%
        \subplot{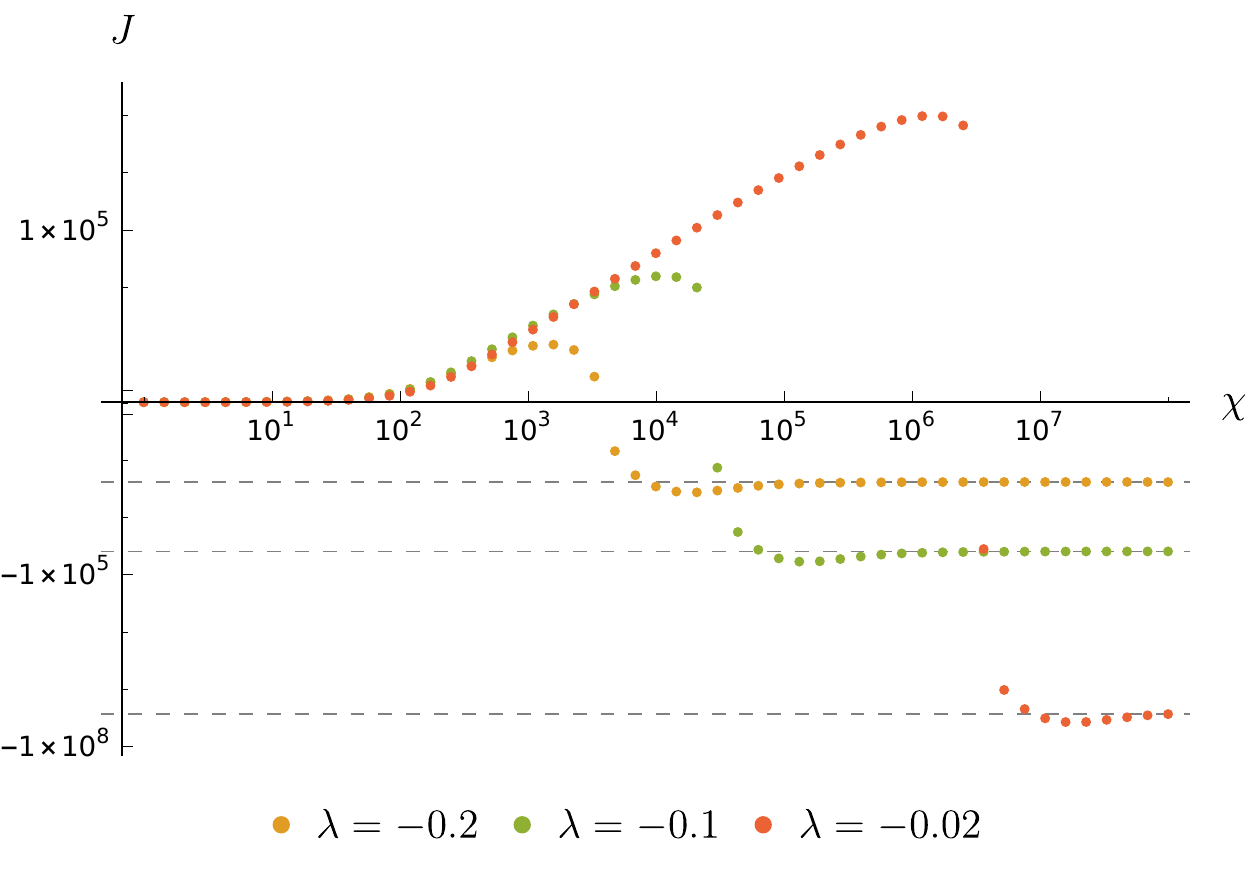}
    }
    \caption{Asymptotic scalings with $\chi$ of integrals $J = \int_0^1 f \, ds$ of the form (\ref{eq:integral-form}) or (\ref{eq:integral-form-t}), cf. subcaptions.
        Each dot represents one numerical integration.
        Lines indicate power law scaling, including asymptotic constants.
        In panels \protect\subref{subfig:ee-neg} and \protect\subref{subfig:kk-neg}, the vertical axis shows $\arcsinh(J/c)$ for a scale $c$; this interpolates symmetrically around $0$ between a linear scale for values $\ll c$ and a logarithmic scale for values $\gg c$, allowing us to show both positive and negative values over a wide range of magnitudes.
        The steepness of the smaller-$\lambda$ curves as the integral changes sign is an artefact of using the same scale $c$ for all curves; see Fig.~\ref{fig:normalised-data}.
        Note that different $\lambda$ lead to very similar curves;
        this is explained by rescaling $t \mapsto t/|\lambda|^3$ in \eqref{eq:airy-exp} and \eqref{eq:contour-integral}, see Appendix~\ref{app:figures} for details.\label{fig:leading-scalings}
    }
\end{figure}

With the preceding results, we can go through each term in the three contributions \eqref{eq:P-ee}--\eqref{eq:P-kk} to the probability, and identify the dominant terms at large $\chi$. These are:
\begin{itemize}
    \item \underline{At order $e^2$:} both $\Ai'$ terms scale like $\chi^{2/3}$.
    Viz.,
    \begin{equation}
    \label{FFS1}
            \int_0^1 \! \ud s\, \frac{1}{z} \Ai'(z) \, 
            = \int_0^\infty\!\ud t\, \frac{\chi t^{-2/3}}{(1 + \chi t)^2} \Ai'(z)
            = O(\chi^{2/3})
    \end{equation}
    according to (\ref{eq:lower-integral-scaling}) and likewise,
    \[
        \int_0^1 \!\ud s \, s \chi \sqrt{z}  \Ai'(z)
        =
        \int_0^\infty\!\ud t\, \frac{\chi^3 t^{4/3}}{(1 + \chi t)^3} \Ai'(z)
        = O(\chi^{2/3})
        .
    \]
    \item \underline{At order $e \kappa$}: the dominant term comes from, assuming $k > 0$,
    \[
            s \chi z^{1/2} \Ai_2(z - k/\sqrt{z})
            = ks \chi \Ai_1(z - k/\sqrt{z}) + \text{subleading}
    \]
    where for small $t$, using the expansion of $\Ai_1$ around $-\infty$,
    \begin{equation}
    \label{FFS2}
        \int_0^\varepsilon \!\ud t\, \frac{\chi^3 t}{(1 + \chi t)^3} \left[
            1 - \frac{t^{1/4}}{\sqrt{\pi}} \cos \left( \frac{2}{3}k^{3/2} t^{-1/2} - \pi/4 \right)
        \right]
        = O(\chi)
    \end{equation}
    after performing the integral with the constant term. (The $\cos$ term is $O(\chi^0)$, as explained above.) We note that this is a faster scaling than in standard QED.
    \item \underline{At order $\kappa^2$}: the dominant term arises from
    \begin{equation}
        \int_0^1\!\ud s\, s \chi z^{-1/2} \Ai(z)
        =
        \int_0^\infty\!\ud t\, \frac{\chi^3 t^{2/3}}{(1 + \chi t)^3} \Ai(z) = O(\chi^{4/3})
        \label{eq:dom-integral}
        \;.
    \end{equation}
    To show this, observe that for $t$ above some cutoff and $\chi$ large enough, we have $\chi t \gg 1$, thus the tail of the integral goes like $\chi^0$.
    Below the cutoff, expanding $\Ai(z)$ as in (\ref{eq:maclaurin}) and performing the integral shows that  it is $O(\chi^{4/3})$, as in~(\ref{eq:lower-integral-scaling}). For sufficiently large $\chi$, this will become the dominant term of the whole probability.
\end{itemize}
Numerical data which confirm the above are presented in Fig.~\ref{fig:leading-scalings} -- the dominant scaling, for large $\chi$, is $\propto \chi^{4/3}$ and comes from the Pauli vertex in the amplitude (mod-squared).

\begin{figure}[t!]
    \centering
    \subfloat{
        \subplot{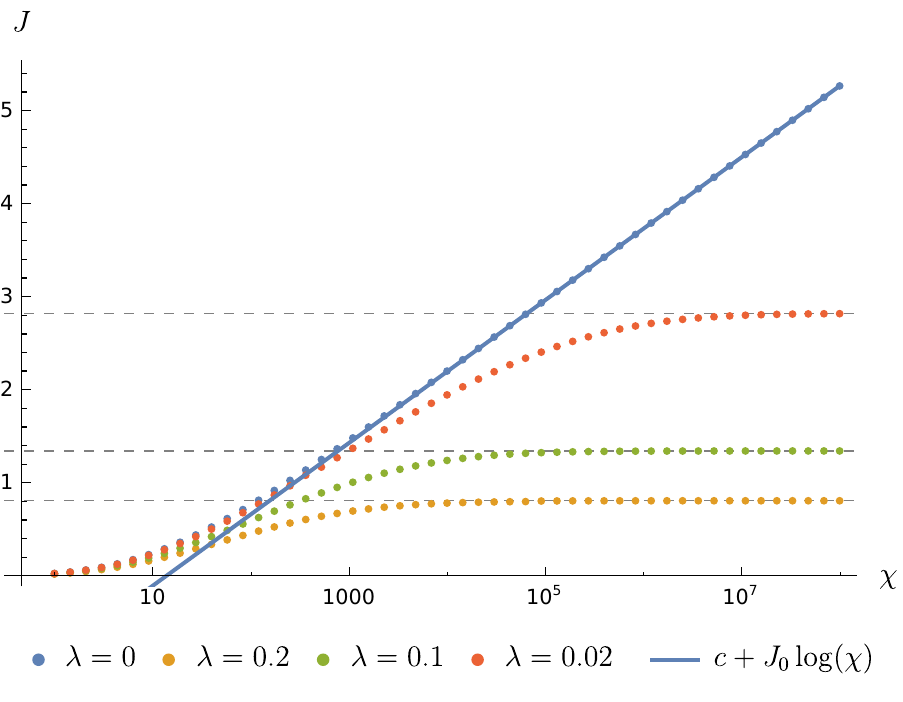}
    }
    \subfloat{
        \subplot{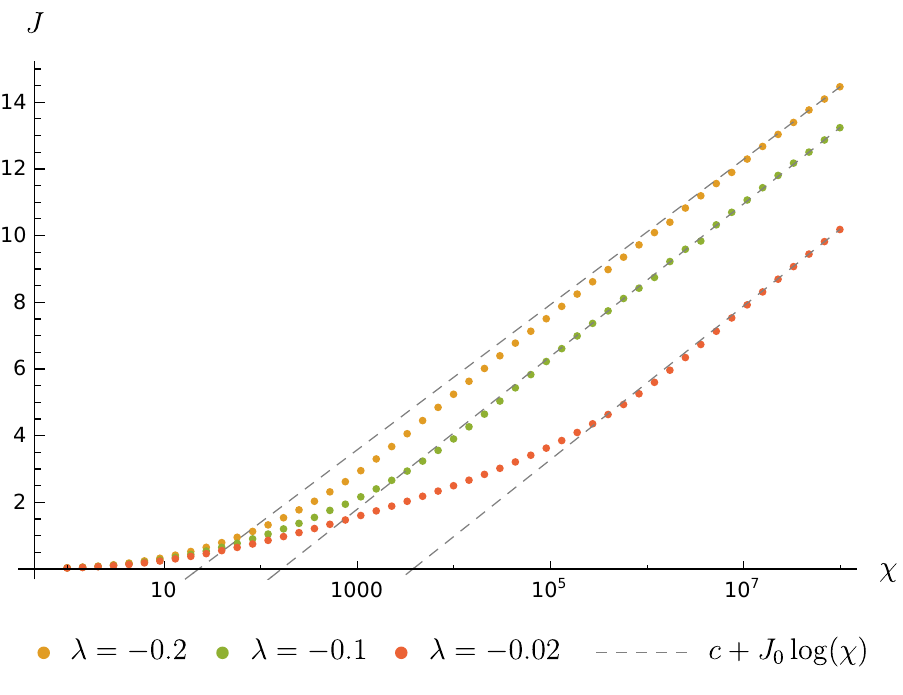}
    }
    \caption{The integral $J = \int_0^1 s \chi z^{3/2} \Ai(z + \lambda/\sqrt{z} )\, ds$ scales as $\log \chi$ for $\lambda \le 0$; naive power counting would suggest $\chi^0$.}
    \label{fig:log-scaling}
\end{figure}

We note that the scaling of the majority of terms \emph{without} shift operators is the same as would be obtained from naive power counting (with $z \sim \chi^{-2/3}$). There are however, from (\ref{eq:lower-integral-scaling}), terms which scale logarithmically, exemplified in Fig.~\ref{fig:log-scaling}, whereas a naive estimate would suggest $O(\chi^0)$.
Further, for terms \emph{with} shift operators, power counting is often incorrect.
An example of this is illustrated in  Fig.~\ref{fig:incorrect-scaling} where the $\lambda = 0$ scaling is $\chi^{4/3}$, but the $\lambda \neq 0$ scaling is $\chi^0$.
The reason naive power counting fails for $\lambda \neq 0$ is that it replaces the Airy function with its value at $0$ when $\chi t \lesssim 1$, but with $\lambda > 0$ this is never the case for large $\chi$, and with $\lambda < 0$ the integrand is, rather, rapidly oscillating around $0$, unless it is the asymptotically constant $\Ai_1$.
The subtleties of the $\lambda \to 0$ limit can be  illustrated by plotting the argument $z + \lambda/\sqrt{z}$ of $\mathcal{A}$ as a function of $z$, Fig.~\ref{fig:airy-argument} in Appendix~\ref{app:figures};
for any non-zero $\lambda$, this function bends away from $z$, such that the limit $\lambda \to 0$ is qualitatively different from any non-zero $\lambda$.

By performing the integral (\ref{eq:dom-integral}), we can work out the coefficient of the dominant ($\chi^{4/3}$ scaling) term in the total probability of nonlinear Compton scattering. 
This gives us the large-$\chi$ dominant rate $\mathbb{R}$ (where $ \mathbb{P} = \int \mathbb{R} \, \ud \varphi $) for nonlinear Compton scattering with a Pauli term,
\begin{equation}
     \mathbb{R}(\kappa \neq 0)
    =
    \frac{\Ai(0) m^2}{ 3^{3/2} n\cdot p} \kappa^2 \chi^{4/3} \approx \frac{0.068 m^2}{n\cdot p} \kappa^2 \chi^{4/3} \;,
\end{equation}
as can be confirmed from the numerical data in Fig.~\ref{fig:incorrect-scaling}, and which can be compared to the usual CCF result~\cite{Ritus:1972ky,Ritus1985a,Fedotov:2016afw},
\begin{equation}
    \mathbb{R}(\kappa = 0)
    =
    -\frac{14 m^2}{3^{ 5/2 } n\cdot p} \Ai'(0) \alpha \chi^{2/3}
    =
    \frac{1.46 m^2}{n\cdot p} \alpha \chi^{2/3}
    \;.
\end{equation}
If we simply equate $\alpha\chi^{2/3} = \kappa^2 \chi^{4/3}$, we see that the dominance of the Pauli term sets in for $\chi$ values of order $(e/\kappa)^3$, or, reinstating the original `new-physics' scale $\Lambda$ from (\ref{eq:action}), $\chi \sim (e/\kappa)^3(\Lambda/m)^3$.
The numerical results also show that subdominant terms, due to the Pauli coupling, remain of magnitude comparable to leading terms for fairly large $\chi$, depending on $\kappa$; see Appendix~\ref{app:figures} for details.
This reflects the interplay between energy and intensity: extremely high intensity is required for the high-energy effects to dominate.

\section{Conclusions}
\label{sec:conclusions}

We have shown that changing the high-energy behaviour of QED induces a change in its high-intensity behaviour.
Due to the presence of the introduced Pauli term, observables such as scattering rates scale differently with intensity compared to ordinary QED.

Considering nonlinear Compton scattering at tree level in Furry picture perturbation theory, we have established the leading-order intensity dependence of the scattering probability (\ref{Prob1}) at high intensity.
Regarding the three contributions (\ref{eq:P-ee})--(\ref{eq:P-kk}) and identifying their dominant high-$\chi$ scaling in (\ref{FFS1})--(\ref{eq:dom-integral}), we see that each Pauli vertex  introduces an additional factor $\chi^{1/3}$ to the high intensity scaling: the dominant terms coming from the QED vertex squared, the interference term, and the Pauli vertex squared, cf.~\eqref{eq:m-sq-expanded}, scale as
\begin{equation}
\label{tre}
    \big(e \chi^{1/3}\big)^2
    \,,
    \qquad
    \big(\kappa \chi^{2/3}\big) \big(e \chi^{1/3} \big) 
    \,, \qquad
    \text{and}
    \qquad
    \big(\kappa \chi^{2/3}\big)^2
\end{equation}
respectively.
Now, there are two ways of viewing high-energy corrections to nonlinear Compton scattering: either through the explicit Pauli term; or through loop effects which, recall, induce a vertex with the same form, $\overline{\psi} \sigma F \psi$. Diagrammatically, we may therefore represent the high-energy corrections as:
    \def\r{1}
    \def\h{0.5}
    \def\q{40}
    \begin{equation}
        \begin{feynman}
            \draw [photon, decoration={number of sines=5}] (0, 0) -- ++ (\q:\r);
            \draw [dressed] (-\r,0) -- (\r, 0);
            \filldraw [fill=white] (0,0) circle (.3);
            \filldraw [fill=white, pattern=north east lines, opacity=1] (0,0) circle (.3);
        \end{feynman}
        =
        \begin{cases}
        \begin{feynman}
            \draw [photon, decoration={number of sines=5}] (0, 0) -- ++ (\q:\r);
            \draw [dressed] (-\r,0) -- (\r, 0);
        \end{feynman}
        +
        \begin{feynman}
            \draw [photon] (-\r/2, 0) arc (-180:0:\r/2);
            \draw [photon, decoration={number of sines=5}] (0, 0) -- ++ (\q:\r);
            \draw [dressed] (-\r,0) -- (\r, 0);
        \end{feynman}
        + \ldots
        \\
        \begin{feynman}
            \draw [photon, decoration={number of sines=5}] (0, 0) -- ++ (\q:\r);
            \draw [dressed] (-\r,0) -- (\r, 0);
        \end{feynman}
        +
        \begin{feynman}
            \draw [photon, decoration={number of sines=5}] (0, 0) -- ++ (\q:\r);
            \draw [dressed] (-\r,0) -- (\r, 0);
            \filldraw [fill=white] (0,0) circle (.1);
        \end{feynman}
        + \ldots
        \end{cases}
    \end{equation}
Mod-squaring the bottom line yields the three terms of \eqref{tre}. In the upper line, the QED one-loop vertex correction in a CCF scales like $e^3 \chi^{2/3}$ at high $\chi$~\cite{Morozov:1981pw}, which reproduces the scaling of the Pauli vertex upon replacing the QED vertex count, $e^3$, by the Pauli coupling $\kappa$. Thus,
\eqref{tre} provides a useful consistency check: both the QED and effective field theory view lead to the same high-$\chi$ behavior. Of course, the effective field theory can only be expected to capture part of the full theory;
for instance, we should not expect that the spectrum in \eqref{tre-termer} agrees in detail with that computed using the loop correction.
It has also been shown that Schwinger pair production is dominated by the $\sigma F$ term of the Dirac operator squared, making a qualitative difference between spinor and scalar QED~\cite{Gies:2016coz}.

As seen in Fig.~\ref{fig:leading-scalings}, the high-intensity power law scaling only sets in for very large values of $\chi$, say for $\chi \sim 10^3$.
At such extreme intensities, backreaction on the strong field is likely to be non-negligible~\cite{Bell:2008zzb,Seipt:2016fyu,Blackburn:2019rfv}, and a treatment beyond the external field approximation~\cite{Ilderton:2017xbj} is necessary.
To date the impact of back-reaction does not seem to have been considered in the context of the Narozhny-Ritus conjecture, and it would be interesting to address this. (For a toy model illustrating that backreaction cannot be neglected see~\cite{Ekman:2020vsc}.)
One could also consider, with the inclusion of a Pauli term, higher orders in perturbation theory, or resummation to all orders as in~\cite{Mironov:2020gbi}.
Given our results, it would also be interesting to revisit the case of non-constant fields, where the asymptotics depend on energy and intensity \emph{separately}~\cite{Ilderton:2019kqp,Podszus:2018hnz}; this clearly ties into the interrelation of high intensity and high energy we have studied here.

\begin{acknowledgements}
\textit{The authors are supported by The Leverhulme Trust, project grant RPG-2019-148.}
\end{acknowledgements}

\appendix

\section{Explicit lightfront helicity spinors}
\label{app:basis}

In the basis where the $\gamma$ matrices take the form
\begin{align}
    \gamma^0 =
    \begin{pmatrix}
        0 & -i\mathbb{I}\\
        i\mathbb{I} & 0
    \end{pmatrix}
    \quad
    \gamma^1 =
    \begin{pmatrix}
        -i\sigma_2 & 0 \\
        0 & i \sigma_2
    \end{pmatrix}
    \quad
    \gamma^2 =
    \begin{pmatrix}
        i\sigma_1 & 0 \\
        0 & -i \sigma_1
    \end{pmatrix}
    \quad
    \gamma^3 =
    \begin{pmatrix}
        0 & i \mathbb{I} \\
        i \mathbb{I} & 0
    \end{pmatrix}
    ,
\end{align}
$\sigma_i$ being the $2 \times 2$ Pauli matrices, the explicit forms of the lightfront helicity eigenspinors are
\begin{equation}
    u_{p \LCp}
    =
    \begin{bmatrix}
        2\sqrt{p_\LCm} \\
        0 \\
        \frac{im}{2 \sqrt{p_\LCm} } \\
        \frac{p_2 - i p_1}{2 \sqrt{p_\LCm} }
    \end{bmatrix}
    \qquad
    u_{p \LCm}
    =
    \begin{bmatrix}
        0 \\
        2\sqrt{p_\LCm} \\
        \frac{p_2 + i p_1}{2 \sqrt{p_\LCm} } \\
        \frac{im}{2 \sqrt{p_\LCm} }
    \end{bmatrix}
    \;,
\end{equation}
normalised to ${\bar u}_{p\LCpm}u_{p\LCpm}=2m$ and ${\bar u}_{p\LCpm}u_{p\LCmp}=0$. Using these results, the relations (\ref{useful}) and (\ref{eq:rot-mat}) in the text are easily verified.

\section{Behaviour of integrals with negative $k$}
\label{app:figures}

\begin{figure}[t!]
    \centering
    \subplot{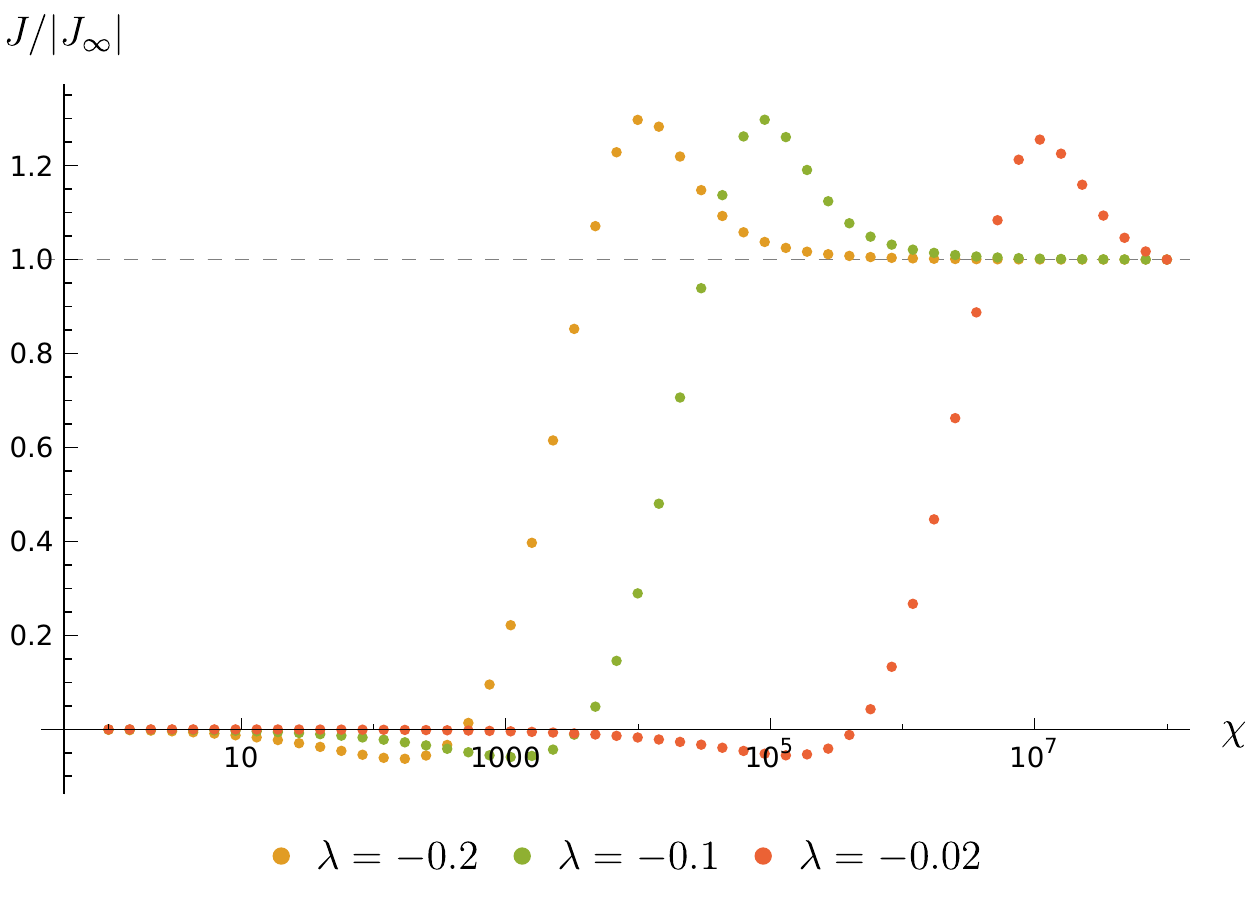}
    \subplot{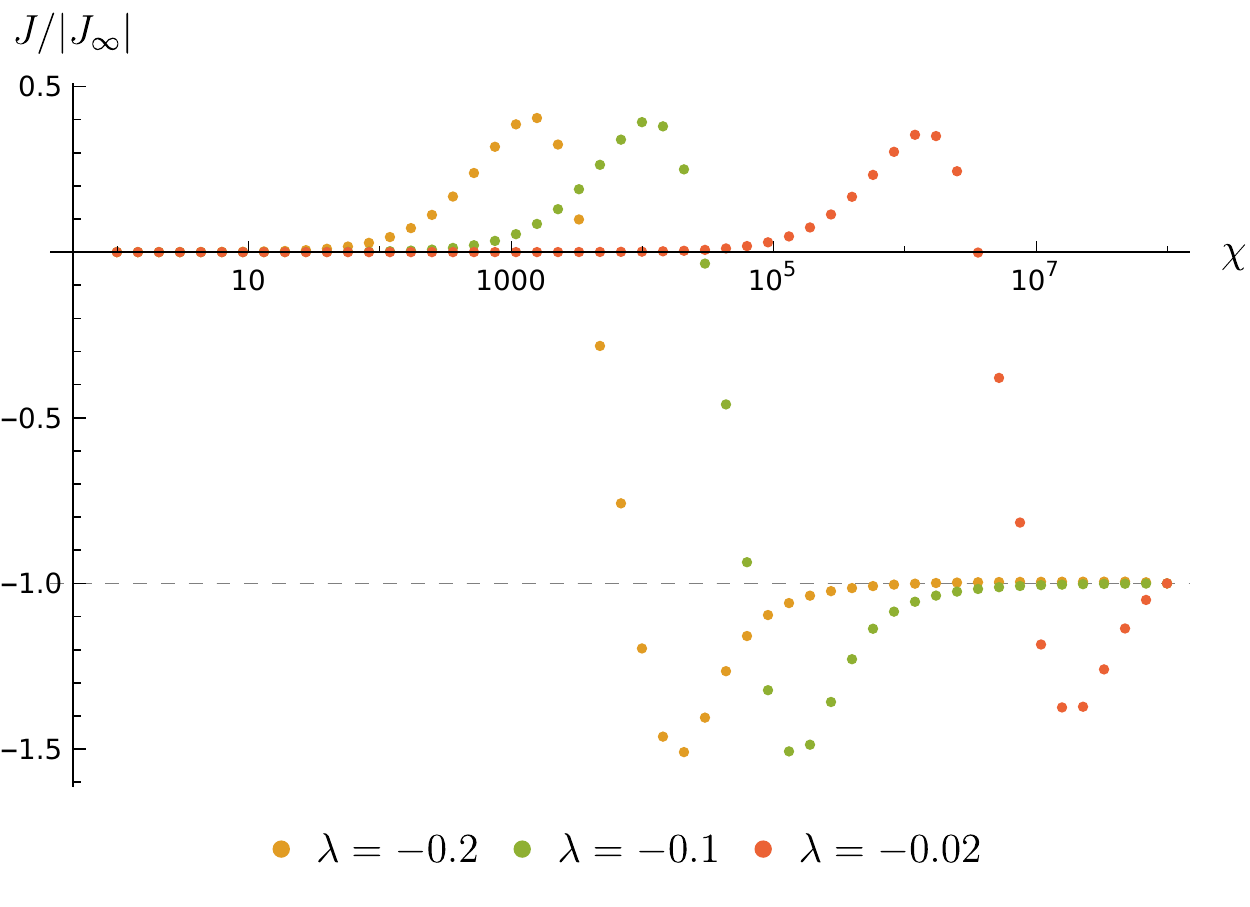}
    \caption{The data from Figs.~\ref{subfig:ee-neg} and \ref{subfig:kk-neg}, normalised to the asymptotic constant value $J_\infty$ for each $\lambda$.}
    \label{fig:normalised-data}
\end{figure}

We show here that the steep swings in Figs.~\ref{subfig:ee-neg} and \ref{subfig:kk-neg} are artefacts of the scale used. In Fig.~\ref{fig:normalised-data} we plot the same data but normalised to the asymptotic values, so that the range of values is similar for each $\lambda$.
The curves are seen to have very similar shapes for all values of $\lambda$, differing only by shifts along the $\chi$-axis.
This can be understood from \eqref{eq:contour-integral}, since rescaling $u \mapsto |\lambda|^{3/2} u $ puts this integral in the form
\begin{equation}
    \sim |\lambda|^{3b' + 3} \int_{\varepsilon^{-1/2}}^\infty \ud u \, \frac{u^{2a - 2b' - 3}}{(u^2 + |\lambda|^3 \chi)^a} \cos(u) 
\end{equation}
such that the integral depends on $\lambda$ only through an overall multiplicative factor, and through the product $|\lambda|^3 \chi$;
this argument also applies to the panels on the left in Fig.~\ref{fig:leading-scalings}.
From the argument following \eqref{eq:contour-integral}, the integral goes like $|\lambda|^{3b' - 3a + 3} \chi^{-a}$ for large $\chi$;
this results in $|\lambda|^{-9/4}$ for Fig.~\ref{subfig:ee-neg} and $|\lambda|^{-15/4}$ for Fig.~\ref{subfig:kk-neg}, explaining the difficulty in displaying all the data on the same scale.
(A log scale could be used for $|J|$, but would have a misleading cusp at the zero-crossing.)

The integral can change sign when $\lambda < 0$ due to the oscillating (trigonometric) behaviour of the Airy function for negative arguments.
The precise mechanism can be made clearer by considering the vertical part of the contour in Fig.~\ref{fig:contour}, where, after rescaling, the denominator is $( \varepsilon^{-1} + 2 i \varepsilon^{-1/2} u - u^2 + |\lambda|^3 \chi )^a$, and the integrand carries an exponential factor $e^{-u}$.
The sign of the integral thus depends on where in the complex plane the denominator falls for $u \lesssim 1$, which depends on the product $|\lambda|^3 \chi$.
Indeed, in Fig.~\ref{fig:normalised-data}, the zero-crossing can be seen to occur for $\chi$ a factor of $10^3$ larger when $|\lambda|$ is a factor of $10$ smaller.

\begin{figure}[t!]
    \centering
    \includegraphics[width=0.4\columnwidth]{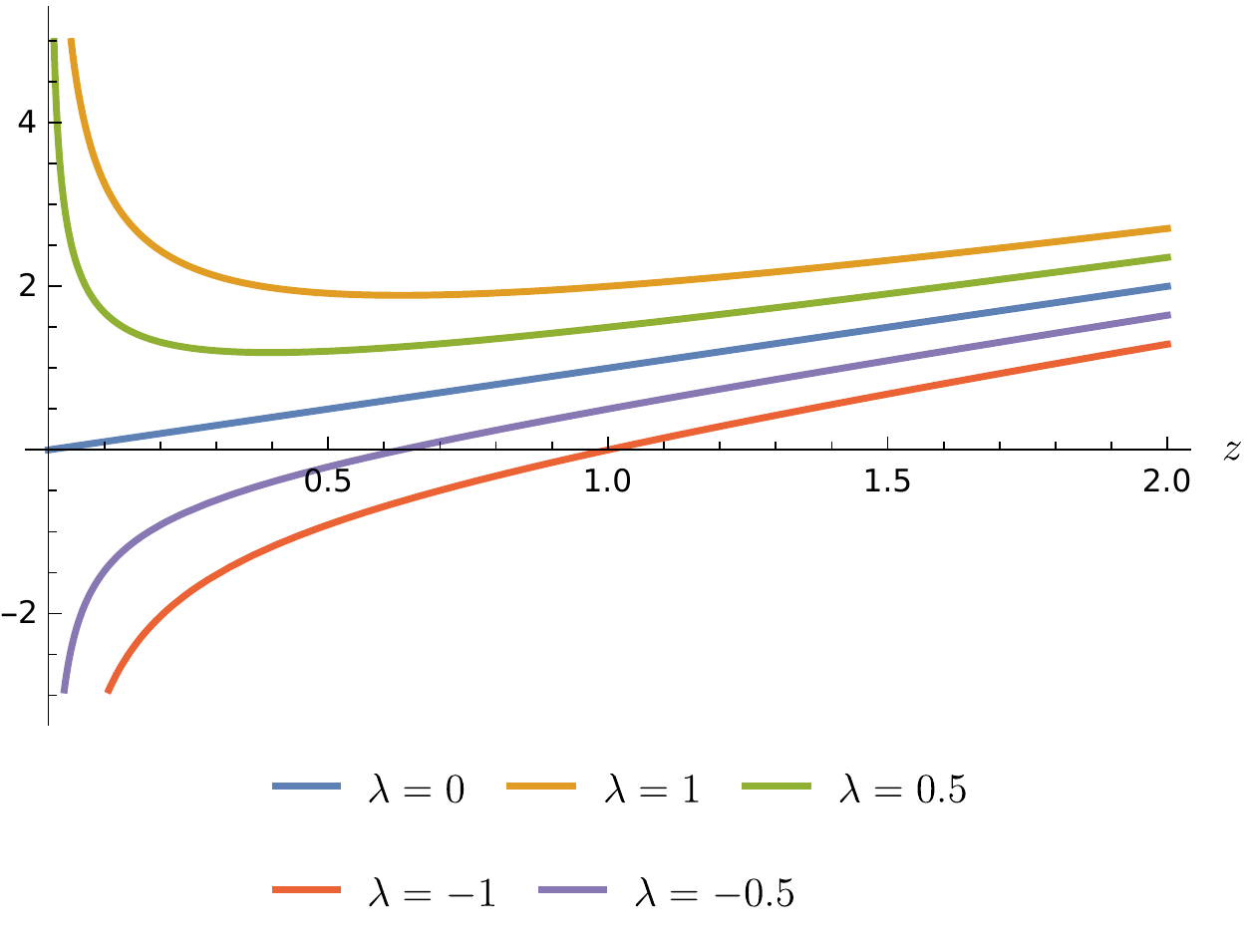}
    \caption{For any $\lambda \neq 0$, the argument of $\mathcal{A}(z + \lambda/\sqrt{z})$ bends away from the straight line for small $z$, making the $\lambda \neq 0$ case qualitatively different from the $\lambda = 0$ case in integrals like (\ref{eq:integral-form}).
    }
    \label{fig:airy-argument}
\end{figure}

\bibliography{PauliBib}

\end{document}